\documentclass[journal,onecolumn,12pt]{IEEEtran}

\usepackage{epsfig,latexsym}
\usepackage{float}
\usepackage{indentfirst}
\usepackage{amsmath}
\usepackage{amssymb}
\usepackage{times}
\usepackage{subfigure}
\usepackage{psfrag}
\usepackage{cite}
\usepackage{lastpage}
\usepackage{fancyhdr}
\usepackage{color}

\newtheorem{Lemma}{Lemma}

\newtheorem{theorem}{$\mathbf{Theorem}$}
\newtheorem{lemma}[Lemma]{$\mathbf{Lemma}$}

\begin{document}
\title{ {   Energy Harvesting Cooperative Networks: Is the Max-Min   Criterion Still Diversity-Optimal? }}

\author{ Zhiguo Ding, \IEEEmembership{Member, IEEE} and  H. Vincent Poor, \IEEEmembership{Fellow, IEEE}\thanks{
Z. Ding and H. V. Poor are with the Department of
Electrical Engineering, Princeton University, Princeton, NJ 08544,
USA.    Z. Ding is also with the School of
Electrical, Electronic, and Computer Engineering, Newcastle
University, NE1 7RU, UK.  }} \maketitle\vspace{-4em}
\begin{abstract}
This paper considers a general energy harvesting cooperative network with $M$ source-destination (SD) pairs and one relay, where the relay   schedules only $m$ user pairs for transmissions. For the special case of $m=1$, the addressed scheduling problem is equivalent to relay selection for the scenario with one SD pair and $M$ relays. In conventional cooperative networks, the max-min selection criterion has been   recognized  as a diversity-optimal strategy for relay selection and user scheduling. The main contribution of this paper is to show  that  the use of the max-min criterion will result in loss of diversity gains in energy harvesting cooperative networks.   Particularly when only a single user is scheduled, analytical results are developed to demonstrate that  the diversity gain achieved by the max-min criterion is only $\frac{M+1}{2}$, much less than the maximal diversity gain $M$. The max-min criterion suffers this diversity loss because it does not reflect the fact that the source-relay channels are more important than the relay-destination channels in energy harvesting networks. Motivated by this fact, a few user scheduling approaches tailored to energy harvesting networks are developed and their performance is analyzed. Simulation results are provided to demonstrate the accuracy of the developed analytical results and facilitate the performance comparison.
 \end{abstract}

\section{Introduction}
Simultaneous wireless information and power transfer (SWIPT) has recently received a lot of attention. Compared to conventional energy harvesting techniques, SWIPT can be used even if wireless nodes do  not have access to external energy sources, such as solar and winder power. 
The key idea of SWIPT is to collect energy from radio frequency (RF) signals, and this new concept of energy harvesting was first proposed in \cite{varshney08} and \cite{Grover10}. Particularly by assuming that the receiver has the capability to carry out energy harvesting and information decoding at the same time, the tradeoff between information rate and harvested energy has been characterized in \cite{varshney08} and \cite{Grover10}. Motivated by the difficulty  of designing  a circuit performing both energy harvesting and signal detection simultaneously, a practical receiver architecture has been developed in \cite{Zhouzhang13}, where two receiver strategies, power splitting and time sharing, have been proposed and their performance have been analyzed.

The concept  of SWIPT was  initially studied in simple scenarios with one source-destination pair, where the use of co-channel interference for energy harvesting was considered in \cite{LiangLiu12} and the combination of multiple-input multiple-output (MIMO) technologies with SWIPT was investigated in \cite{Xiangzt12}. SWIPT has  been recently applied to various important communication scenarios more complicated than the case  with one source-destination pair. For example, in \cite{Juzhangh13} the application of SWIPT to multiple access channels has been considered, where a few solutions for system throughput maximization have been proposed.   Broadcasting scenarios have   been considered in \cite{Ruizhangbroadcast13} and \cite{Huangl13},  where one transmitter is to  serve two types of users, energy receivers and information receivers, simultaneously. In \cite{Liuzhangmiso2014} the joint design of uplink information transfer and downlink energy transfer has been  considered, where sophisticated algorithms  for energy beamforming, power allocation and throughput maximization have been proposed. The idea of SWIPT has also been applied to wireless cognitive radio systems, where opportunistic energy harvesting from RF signals has been studied in \cite{Ruicogswipt}.

The application of SWIPT to cooperative networks is important since the lifetime of the relay batteries can be extended by efficiently using the energy harvested from the relay observations. In \cite{Krikidis12} a greedy switching approach between data decoding and energy harvesting has been  proposed for the case with one source-destination pair and one relay. In \cite{Nasirzhou} the outage performance achieved by amplified-and-forward (AF) relaying protocols has been developed, and the use of decode-and-forward (DF) strategies has been investigated  in multi-user energy harvesting cooperative networks \cite{Dingpoor133}. Relay selection has been studied in a broadcasting scenario where energy harvesting was carried out at the destinations, instead of relays \cite{Himal141}.  The impact of the random locations of wireless nodes on the path loss and the outage performance has been characterized by  applying stochastic geometry in \cite{Dingpoor132}.

In conventional cooperative networks, the max-min criterion has been recognized as a diversity-optimal selection strategy  \cite{Krikidis091,Jingmaxmin1,Song11maxmin}. Take a DF cooperative network with one source-destination pair and $M$ relays as an example. Provided that the $i$-th relay is used, the capacity of a DF relay channel is $\min\{\log(1+\rho |h_i|^2), \log(1+\rho |g_i|^2)\}$, where $\rho$ is the transmit signal-to-noise ratio (SNR), $h_i$ is the channel gain between the source and the relay, and $g_i$ is the channel gain between the relay and the destination. Obviously the max-min criterion, i.e. $\max\{\min\{|h_i|^2,|g_i|^2\}, 1\leq i \leq M\}$, is capacity optimal and can achieve the maximal diversity gain, $M$. But is this conclusion still valid when   energy harvesting relays are used?

The main contribution of this paper is to characterize the performance of the max-min selection criterion in energy harvesting cooperative networks. We first construct a general framework of energy harvesting cooperative networks, where $M$ pairs of sources and destinations communicate with each other via a relay. Among the $M$ user pairs, the relay will schedule $m$ of them to transmit. It is important to point out that the problem of relay selection for the scenario with one source-destination pair is a special case of the formulated framework by setting $m=1$. When only a single  user is scheduled, the exact expression for the outage probability achieved by the max-min criterion is developed by carefully grouping the possible outage events  and then applying order statistics. Based on this obtained expression, asymptotic studies of the outage probability are carried out to show that the diversity gain achieved by the max-min criterion is only $\frac{M+1}{2}$, much less than the full diversity gain, $M$.

The reason for this loss of the diversity gain is that the max-min criterion treats the source-relay channels and the relay-destination  channels equally. However, when an energy harvesting relay is used, it is important to observe that the source-relay channels become more important. For example, the source-relay channels   impact not only the reception reliability at the relay, but also the relay transmission power. Recognizing this fact, a few modified user scheduling approaches are developed, which is the second contribution of this paper. Particularly for the case of $m=1$,  an efficient  user scheduling approach is proposed, and analytical results are developed to demonstrate that this approach can achieve the maximal diversity gain. This approach can be extended to the case of $m>1$, by applying exhaustive search. A greedy user scheduling approach is also developed by assuming that the relay always has data to be sent to all the destinations. The use of this greedy approach   yields closed-form expressions for the outage probability and diversity order, which can be used as an upper bound for the other approaches. Simulation results are also provided to demonstrate the accuracy of the developed analytical results and facilitate the performance comparison among the addressed user scheduling approaches.


\section{System Model}\label{section II}
Consider a cooperative communication scenario with $M$ source-destination pairs and one {\it energy harvesting} relay. The $M$ users compete  for the wireless medium, and the relay will schedule $m$ user pairs over $2m$ time slots, $0\leq m \leq M$.  All the channels are assumed to be independent and identically (i.i.d.) quasi-static Rayleigh fading, and this indoor slow fading model is   valid for many applications of wireless energy transfer, such as wireless body area networks and smart homes \cite{Himal141} and \cite{Dingpoor132}. In Section \ref{section simulation}, the impact of the path loss and the random locations of the users on the outage performance will be studied\footnote{    Note that when the users are randomly deployed, the effective channel gains, i.e. the combinations of Rayleigh fading and large scale path loss,  can be   still approximated as independent and identically  exponentially distributed variables \cite{Dingpoor132}.}. It is assumed that  the relay has   access to   global channel state information (CSI), which is important for the relay to carry out user scheduling.

During the $j$-th time slot, consider that the $i$-th user pair is scheduled to transmit its message $s_i$, where the details for user scheduling will be provided in the next two sections. The power splitting strategy will be used at the DF relay. Particularly the relay will first direct the observation flow to the detection circuit, and then to the energy harvesting circuit if there is any energy left after successful detection \cite{Zhouzhang13} and \cite{Dingpoor133}. Therefore the observation at the relay is given by
\begin{eqnarray}
y_{ri} = \sqrt{P(1-\theta_i)}h_is_i+n_{ri},
\end{eqnarray}
where $\theta_i$ is the power splitting factor, $P$ is the transmission power at the source, $h_i$ denotes the channel gain from the $i$-th source to the relay, and $n_{ri}$ denotes the additive white Gaussian noise. As discussed in \cite{Dingpoor133}, the optimal value of $\theta_i$ for a DF relay is $\max\left\{1-\frac{\epsilon}{|h_i|^2},0\right\}$,   the maximal value of $\theta_i$ constrained by successful detection at the relay, where $\epsilon=\frac{2^{2R}-1}{P}$ and $R$ denotes the targeted data rate. The power obtained at the relay after carrying out energy harvesting from the $i$-th user pair is given by
\begin{equation}
P_{ri} =  \eta P\left[|h_i|^2-\epsilon\right]^+,
\end{equation}
where $\eta$ denotes the energy harvesting coefficient, and $[x]^+$ denotes $\max\{x,0\}$. At the $(m+j)$-th time slot, the relay forwards $s_i$ to the $i$-th destination, and the receive SNR at this destination is given by
\begin{eqnarray}
SNR_i = P_{i} |g_i|^2,
\end{eqnarray}
where $P_{i}$ denotes the relay transmission power allocated to the $i$-th destination, and $g_i$ denotes the channel gain between the relay and the $i$-th destination. Note that $P_{ri}$ is not necessarily equal to $P_i$, depending on the used relay strategy, as discussed in the following sections.

\section{The Performance Achieved by The Max-min Criterion}\label{section III}
\subsection{User scheduling  based on the max-min criterion}
 In this section, the performance achieved by the user scheduling strategy based on the max-min criterion is studied. Particularly we will focus on the case that the relay selects only one user pair, i.e. $m=1$, and more discussions about the case with $m>1$ will be provided in the next section.   Note that the scenario addressed in this section  can be shown mathematically the same as the problem of relay selection for the case with one source-destination pair  and $M$ relays. Therefore the results obtained for the addressed scheduling problem  will be also applicable to the max-min relay selection cases.

Since only one user pair is scheduled, the energy harvested from the $i$-th source will be used to power the relay transmission to the $i$-th destination,   i.e. $P_{i}=P_{ri}$.  The max-min   user scheduling strategy can be described as follows:
\begin{itemize}
\item The relay first finds out the worst link of each user pair. Denote $z_i= \min\{|h_i|^2, |g_i|^2\}$.
\item The user pair with the strongest worst link is selected, i.e. the $i^*$-th user pair is selected because $i^*=\arg \max \left\{z_1, \ldots, z_M\right\}$.
\end{itemize}

Provided that the relay can decode the $i^*$-th source's message correctly, the SNR at the corresponding destination is given by
\begin{eqnarray}
SNR_{i^*} = \eta P\left(|h_{i^*}|^2-\epsilon\right) |g_{i^*}|^2.
\end{eqnarray}

\subsection{Performance evaluation}
The outage probability achieved by the max-min based scheduling  scheme  can be written as follows:
\begin{eqnarray}\label{outage probability}
\mathrm{P}_{o}\triangleq\mathrm{P}\left(|h_{i^*}|^2<\epsilon\right) + \mathrm{P}\left((|h_{i^*}|^2-\epsilon)|g_{i^*}|^2<\epsilon_1, |h_{i^*}|^2>\epsilon \right),
\end{eqnarray}
where $\epsilon_1=\frac{\epsilon}{\eta}$.  Although the outage probability achieved by the max-min criterion is shown in a simple term as in \eqref{outage probability}, it is challenging to evaluate this probability. The reason is that the use of the   scheduling strategy has changed the statistical property of the channel gains. For example, $|h_{i^*}|^2$ is no longer exponentially distributed. The density function of $\min\{|h_{i^*}|^2,|g_{i^*}|^2\}$ can be found by using order statistics, and the key step is to restructure the expression of the outage probability shown in \eqref{outage probability} into a form to which the density function of  $\min\{|h_{i^*}|^2,|g_{i^*}|^2\}$ can be applied. In the following theorem, the exact expression for the outage probability achieved by the max-min scheme is provided.
\begin{theorem}\label{theorm1}
When a single user is scheduled, the outage probability achieved by the max-min   user scheduling strategy is given by
\begin{eqnarray}\label{theorem eq}
\mathrm{P}_o&=& \frac{e^{-\epsilon}}{2}\sum_{i=0}^{M}{M \choose i} \frac{(-1)^i}{2i-1}\left(1-e^{-(2i-1)\epsilon}\right) \\\nonumber &&+M \sum^{M-1}_{i=0}{M-1 \choose i} (-1)^i  \left(\frac{e^{-\epsilon}-e^{-(2i+2)\epsilon}}{2i+1} +\frac{e^{-(2i+2)\epsilon}-e^{-(2i+2)\epsilon_0}}{2i+2} - e^{-\epsilon} \beta(\epsilon_0,i) \right)
\\ \nonumber &&+  \frac{\left(1-e^{-2\epsilon}\right)^M}{2}+ M\sum^{M-1}_{i=0} {M-1\choose i} (-1)^i \left(\frac{e^{-2(i+1)\epsilon}-e^{-2(i+1)\epsilon_0}}{2(i+1)} -  e^{-(2i+1)\epsilon}\beta(\epsilon_0-\epsilon, i) \right),
\end{eqnarray}
where $\beta(y,i)=\int^{\epsilon_0}_{0}e^{-(2i+1)y-\frac{\epsilon_1}{y}} dy$,  and $\epsilon_0=\frac{\epsilon+\sqrt{\epsilon^2+4\epsilon_1}}{2}$.
\end{theorem}
 \begin{proof}
 See the appendix.
\end{proof}
The   expression shown in \eqref{theorem eq} can be   used to numerically evaluate the outage probability achieved by the max-min scheduling approach, as shown in Section \ref{section simulation}. In addition, it can also be used for the analysis of the diversity gain achieved by the max-min approach, as shown in the following theorem.
\begin{theorem}\label{theorem 2}
When a single   user pair is scheduled, the diversity order achieved by the max-min   user scheduling approach is $\frac{M+1}{2}$.
\end{theorem}
 \begin{proof}
 See the appendix.
\end{proof}
For the addressed topology, there are $M$ independent pathes given $M$ user pairs, which means that the maximal diversity gain is $M$. And Theorem \ref{theorem 2} indicates that the max-min scheduling approach cannot achieve this maximum  diversity.
As a benchmark scheme, recall a conventional cooperative network that has the same topology as the one described in Section \ref{section II}. Without loss generality, let $P_i=P$, i.e. the relay transmission power is the same as the source power.  It can be easily verified that the max-min approach can achieve  the optimal diversity gain, $M$, as shown in the following. The outage probability achieved by the max-min approach is
\begin{eqnarray}\nonumber
\mathrm{P}_o&=&\mathrm{P}(|h_{i^*}|^2<\epsilon)+\mathrm{P}(|g_{i^*}|^2<\epsilon,|h_{i^*}|^2>\epsilon) \\ \nonumber
&=&\mathrm{P}(|h_{i^*}|^2<\epsilon,|g_{i^*}|^2>\epsilon)+\mathrm{P}(|h_{i^*}|^2<\epsilon,|g_{i^*}|^2<\epsilon)
+\mathrm{P}(|g_{i^*}|^2<\epsilon,|h_{i^*}|^2>\epsilon)\\   &=& \mathrm{P}(\min\{|h_{i^*}|^2,|g_{i^*}|^2\}<\epsilon)\rightarrow \epsilon^M,\label{maxmin conventional}
\end{eqnarray}
where the last step is obtained by using the probability density function (pdf)  shown in \eqref{short1} and applying the high SNR approximation.  Comparing \eqref{maxmin conventional} to \eqref{theorem 2}, one can observe that the performance of the max-min scheduling approach in two system setups is significantly different, and  new efficient user scheduling strategies are needed for energy harvesting cooperative networks.

\section{Modified User Scheduling Strategics }\label{section IV}
\subsection{Scheduling a single user pair}\label{subsection 1}
A straightforward   approach of user scheduling for  the energy harvesting scenario  is described as follows:
\begin{itemize}
\item Construct a subset of user pairs containing all the destinations whose source information can be decoded correctly at the relay. Denote this subset as $\mathcal{S}\triangleq \{i\in \mathcal{S}: |h_i|^2\geq \epsilon\}$.
\item  Select a destination from $\mathcal{S}$ to minimize the outage probability of the relay transmission. Denote the index of the selected user by $i^*$, i.e. $i^*=\arg \max \{(|h_i|^2-\epsilon)|g_i|^2, i\in \mathcal{S}\}$.
\end{itemize}

The outage probability achieved by this user scheduling strategy can be expressed as follows:
\begin{eqnarray}
\mathrm{P}_o &\triangleq& \mathrm{P}\left( |S|=0  \right)+\mathrm{P}\left(     (|h_{i^*}|^2-\epsilon)|g_{i^*}|^2 <\epsilon_1,|\mathcal{S}|>0\right) \\ \nonumber &=&
\mathrm{P}\left( |S|=0  \right)+\sum^{M}_{n=1}\underset{T_1}{\underbrace{\mathrm{P}\left(     (|h_{i^*}|^2-\epsilon)|g_{i^*}|^2 <\epsilon_1||\mathcal{S}|=n\right)}}\mathrm{P}(|\mathcal{S}|=n),
\end{eqnarray}
where $|\mathcal{S}|$ denotes the cardinality of the set.
Denote $x_i=|h_i|^2$, and order $x_i$ as $x_{(1)}\leq \cdots \leq x_{(M)}$. The probability of $\mathrm{P}(|\mathcal{S}|=n)$ can be calculated as follows:
\begin{eqnarray}\label{sn}
\mathrm{P}(|\mathcal{S}|=n) &=& \mathrm{P}(x_{(M-n)}<\epsilon, x_{(M-n+1)}>\epsilon)\\ \nonumber &=&   \frac{M!}{(M-n)!n!} \left(1-e^{-\epsilon}\right)^{M-n} e^{-n\epsilon},
\end{eqnarray}
for $0\leq n \leq M$, where the last equation is obtained by applying the joint pdf of $x_{(M-n)}$ and $x_{(N-n+1)}$ \cite{David03} and \cite{Dingtcom07}.   On the other hand $T_1$ can be simply expressed as follows:
\begin{eqnarray}
T_1 &=& \left[\mathrm{P}\left(     (x_i-\epsilon)y_i <\epsilon_1, |i\in \mathcal{S}, |\mathcal{S}|=n\right)\right]^{n},
\end{eqnarray}
where $y_i=|g_i|^2$. In the following we first consider the case of $n\geq 1$. The conditions of $T_1$, $i\in\mathcal{S}$ and $ |\mathcal{S}|=n$, imply   $x\geq \epsilon$, which means that the conditional CDF of $x_i$  is given by
\begin{eqnarray}
F_{x_i|i\in\mathcal{S}, |\mathcal{S}|\geq 1}(x) = \frac{e^{-\epsilon }-e^{-x}}{e^{-\epsilon}},
\end{eqnarray}
 for $x\geq\epsilon$. The two conditions,  $i\in\mathcal{S}$ and $|\mathcal{S}|=n$,  do  not affect  $y_i$ which is still exponentially distributed.  Therefore the factor $T_1$ can be calculated as follows:
\begin{eqnarray}
T_1 &=&  \left(\mathcal{E}_{y}\left(\frac{e^{-\epsilon }-e^{-\frac{\epsilon_1}{y}-\epsilon}}{e^{-\epsilon}}\right)\right)^n\\ \nonumber  &=& \left(1 - 2\sqrt{\epsilon_1}\mathbf{K}_{1}\left(2\sqrt{\epsilon_1}\right)\right)^n,
\end{eqnarray}
where $\mathbf{K}_n(\cdots)$ denotes the modified Bessel function of the second kind.
Recall that $x\mathbf{K}_1(x)\approx 1+\frac{x^2}{2}\ln \frac{x}{2}$ for $x\rightarrow 0$, \cite{Dingpoor133}, which means $T_1\approx \epsilon \ln \frac{1}{\epsilon}$. The overall outage probability can be approximated as follows:
\begin{eqnarray}\label{eqlemma1}
\mathrm{P}_o  &=&
\left(1-e^{-\epsilon}\right)^M +\sum^{M}_{n=1}\left(1 - 2\sqrt{\epsilon_1}\mathbf{K}_{1}\left(2\sqrt{\epsilon_1}\right)\right)^n \frac{M!}{(M-n)!n!} \left(1-e^{-\epsilon}\right)^{M-n} e^{-n\epsilon}\\ \nonumber &\approx& \epsilon^M +\sum^{M}_{n=1} \epsilon^n \left(\ln \frac{1}{\epsilon}\right)^n \frac{M!}{(M-n)!n!} \epsilon^{M-n}.
\end{eqnarray}
When $\epsilon \rightarrow 0$, it is straightforward to show  $\frac{\log \mathrm{P}_o}{\log \epsilon}\rightarrow M$, which results in the following lemma.
\begin{lemma}\label{lemma 1}
The proposed user scheduling strategy   can achieve the full diversity gain $M$.
\end{lemma}
Compared to the maxi-min based approach, the proposed scheduling strategy    can achieve   a larger diversity gain. The reason for this performance improvement is that   the source-relay channels have been given a more important role for use scheduling, compared to the relay-destination channels. Particularly the source-relay channels have been considered when forming $\mathcal{S}$ and also selecting the best user from the set, whereas the relay-destination channels  affect only the second step.

\subsection{Scheduling $m$ user pairs}\label{subsection 3}
The approach proposed in the previous subsection can be extended to the case of scheduling $m$ user pairs, as described in the following.
\begin{itemize}
\item Construct a subset of user pairs, $\mathcal{S}$, as defined in Section \ref{subsection 1}.
 \item Find all possible combinations of the users in  $ \mathcal{S} $,  denoted by $\{\pi_1, \cdots, \pi_{{|\mathcal{S}| \choose \min\{m,|\mathcal{S}|\}}}\}$, where each set contains $\min\{m,|\mathcal{S}|\}$ users, i.e. $\pi_i=\{\pi_i(1),\ldots, \pi_i(\min\{m,|\mathcal{S}|\})\}$.
\item For each possible combination, $\pi_i$, $1\leq i \leq {|\mathcal{S}| \choose \min\{m,|\mathcal{S}|\}}$
\begin{itemize}
    \item  Calculate the accumulated power obtained from energy harvesting, $\sum_{j=1}^{\min\{m,|\mathcal{S}|\}}P_{r\pi_i(j)}$.
    \item Distribute the overall power among $m$ destinations equally, i.e. $P_i=\frac{\sum_{j=1}^{\min\{m,|\mathcal{S}|\}}P_{r\pi_i(j)}}{\min\{m,|\mathcal{S}|\}}$.
    \item Find the worst  outage performance among the $\min\{m,|\mathcal{S}|\}$ users in $\pi_i$, denoted by $\mathrm{P}_{o,\pi_i}$.
    \end{itemize}
\item Select the combination which minimize  the worst user outage performance, i.e. $i^*=\arg \min \{\mathrm{P}_{o,\pi_1}, \cdots, \mathrm{P}_{o,\pi_{{|\mathcal{S}| \choose \min\{m,|\mathcal{S}|\}}}}\}$.
\end{itemize}
This scheduling approach is to exhaustively search all possible combinations of the $|\mathcal{S}|$ user pairs, and one combination will be selected if it can minimize the outage probability  for  the worst user case. Provided that there is a large number of users to be scheduled, the complexity of this exhaustive search scheme can be infeasible due to the  large number of the possible combinations.  Note that in this paper, we consider only the equal power allocation strategy, whereas other power allocation strategies, such as the sequential water filling scheme proposed in \cite{Dingpoor133}, can also be applied.

It is difficult to analyze the performance achieved by the exhaustive search approach, since the channel gains from different combinations might be correlated. Instead, we will propose a greedy approach which is applicable to delay tolerant networks, and also serves as an upper bound for the system performance.
\subsection{Greedy user scheduling approach}\label{subsection 2}
First order all the source-relay channels and the relay-destination channels, i.e. $|h_{(1)}|^2\leq \ldots \leq |h_{(M)}|^2$  and  $|g_{(1)}|^2\leq \ldots \leq |g_{(M)}|^2$. The greedy user scheduling approach can be described as follows:
\begin{itemize}
\item Construct  a subset of user pairs, $\mathcal{S}$, as defined in Section \ref{subsection 1}.
 \item Schedule $\min\{m,|\mathcal{S}|\}$ sources   with the best   source-relay channel conditions   during the first $\min\{m,|\mathcal{S}|\}$ time slots, i.e. the $\min\{m,|\mathcal{S}|\}$ sources with the following channels, $|h_{(M-\min\{m,|\mathcal{S}|\}+1)}|^2\leq \ldots \leq |h_{(M)}|^2$.
\item Calculate the accumulated power obtained from energy harvesting, $\sum_{j=1}^{\min\{m,|\mathcal{S}|\}}P_{r (M-j+1)}$.
    \item Schedule $\min\{m,|\mathcal{S}|\}$ destinations with the best relay-destination channel conditions  during the second $\min\{m,|\mathcal{S}|\}$ time slots, i.e. the $\min\{m,|\mathcal{S}|\}$ destinations with the following channels, $|g_{(M-\min\{m,|\mathcal{S}|\}+1)}|^2\leq \ldots \leq |g_{(M)}|^2$, with equally allocated transmission power, denoted by $P_{\min\{m,|\mathcal{S}|\}}=\frac{\sum_{j=1}^{\min\{m,|\mathcal{S}|\}}P_{r (M-j+1)}}{\min\{m,|\mathcal{S}|\}}$.
\end{itemize}
Note that the scheduled destinations are not necessarily the partners of the scheduled sources, so this greedy approach assumes that the relay always has  data to be transmitted to all the destinations.

Based on the above strategy description, the outage probability at the $i$-th best destination, $1\leq i \leq \min\{m, |\mathcal{S}|\}$, can be written as follows:
\begin{eqnarray}
\mathrm{P}_{oi}&\triangleq& \mathrm{P}\left(|\mathcal{S}|=0\right) +\sum^{M}_{n=1}\mathrm{P}\left(\left.P_{\min\{m,|\mathcal{S}|\}}|g_{(M-i+1)}|^2<(2^{2R}-1)\right||\mathcal{S}|=n\right) \mathrm{P}\left(|\mathcal{S}|=n\right) .
\end{eqnarray}
And the following lemma provides the exact expression of the above outage probability.
\begin{lemma}\label{lemma 2}
The outage probability achieved by the greedy user scheduling approach is given by:
\begin{eqnarray}\label{eqlemma2}
\mathrm{P}_{oi}&\triangleq& \mathrm{P}\left(|\mathcal{S}|=0\right) +\sum^{m}_{n=1} T_2\mathrm{P}\left(|\mathcal{S}|=n\right)+\sum^{M}_{n=m+1}T_3\mathrm{P}\left(|\mathcal{S}|=n\right),
\end{eqnarray}
where $\mathrm{P}(|\mathcal{S}|=n)$ is defined in \eqref{sn}, $T_2= i{M \choose i}\sum^{M-i}_{k=0} {M-i \choose k} \frac{ (-1)^k}{k+i}  \left(1-  \frac{2\left((k+i)n\epsilon_1\right)^{\frac{n}{2}}}{(n-1)!}\mathbf{K}_n\left(2\sqrt{(k+i)n\epsilon_1}\right) \right)$, $T_3=\frac{M!}{(M-i)!(i-1)!}\sum^{M-i}_{l=0}{M-i \choose l} \frac{(-1)^l}{l+i} \left( 1- T_4\right)$, $T_4=\sum^{n-m-1}_{k=0} d_{m,k}\left(\sum^{m}_{j=1}\frac{2a_{j,k}\left(m\epsilon_1(l+i)\right)^{\frac{j}{2}} \mathbf{K}_{j}\left(2\sqrt{m\epsilon_1(l+i)}\right)   }{(j-1)!} \right.$
$ \left. +2b_k \sqrt{\frac{m\epsilon_1(l+i)}{1+\frac{k+1}{m}}} \mathbf{K}_1\left(2\sqrt{\epsilon_1(l+i)\left(m+ k+1 \right)}\right)\right)$,  $d_{m,k}=\frac{n!}{(n-m-1)!m!m}{n-m-1 \choose k} (-1)^k$, $b_k=(-1)^m\frac{m^{m}}{(k+1)^{m}}$, and $a_{j,k}= \frac{(-1)^{m-j}m^{m-j+1}}{(k+1)^{m-j+1}}$.
\end{lemma}
 \begin{proof}
 See the appendix.
\end{proof}
Although the outage probability expression in Lemma \ref{lemma 2} can be used for numerical studies, this form is quite complicated and cannot be used for analyzing  diversity gains. For the special case of $m=1$, asymptotic studies can be carried out and the achievable diversity gain can be obtained, as shown in the following lemma.
\begin{lemma}\label{lemma 3}
When scheduling only a single user pair, i.e. $m=1$, the diversity gain achieved by the greedy user scheduling approach is $M$.
\end{lemma}
 \begin{proof}
 See the appendix.
\end{proof}

The fact that   the greedy user scheduling approach can achieve the full diversity gain is not surprising, since the greedy approach outperforms the diversity-optimal one described in Section \ref{subsection 1}.
\section{Numerical Results}\label{section simulation}
In this section, computer simulations will be carried out to evaluate the performance of the user scheduling approaches addressed in this paper. To simplify clarifications, we term the user scheduling approaches described in Section \ref{subsection 1}, \ref{subsection 3} and \ref{subsection 2} as ``Approach I", ``Approach II", and ``Approach III", respectively.

\begin{figure}[!htbp]\centering
    \epsfig{file=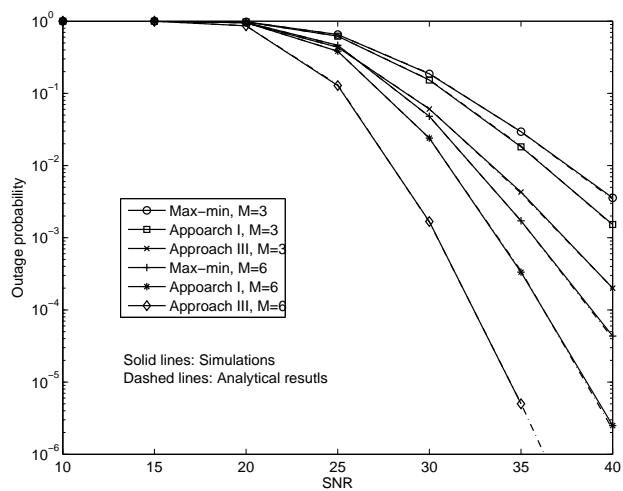, width=0.48\textwidth, clip=}
\caption{Analytical results vs computer simulations. Only one user pair is scheduled, $\eta=1$. The targeted data rate is $R=4$ BPCU. }\label{fig_1}\vspace{-1em}
\end{figure}

\begin{figure}[!htbp]\centering
    \epsfig{file=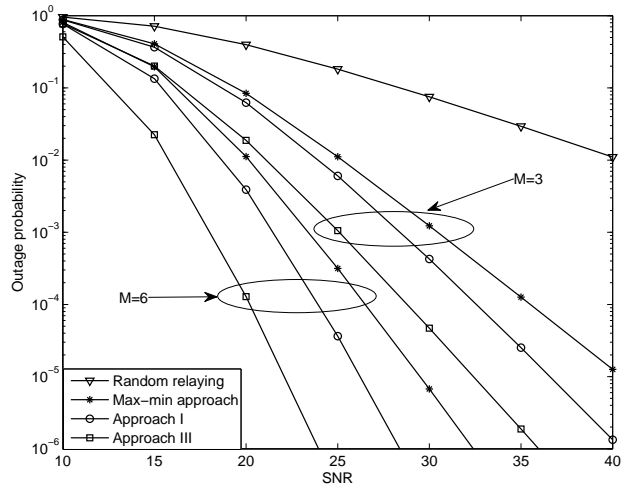, width=0.48\textwidth, clip=}
\caption{Comparison of various user scheduling approaches. Only one user pair is scheduled.  $\eta=1$. The targeted data rate is $R=2$ BPCU. }\label{fig_2}\vspace{-1em}
\end{figure}

We first focus on the scenario where only a single user is scheduled. In Fig. \ref{fig_1} the accuracy of the developed analytical results about the outage probability shown in Theorem \ref{theorm1}, \eqref{eqlemma1}, and Lemma \ref{lemma 2}, is verified by using simulation results, where the targeted data rate is $R=4$ bits per channel use (BPCU), and the energy harvesting efficiency coefficient is $\eta=1$. As can be seen from the figure, the developed analytical results match the simulation results exactly. In Fig. \ref{fig_2} the outage probabilities achieved by different user scheduling approaches are examined with more details, where analytical results are used to generate the figure. As a benchmark, the scheme with a random selected user   is also shown in the figure, and its outage performance is the worst among all the scheduling approaches.   On the other hand, Approach III, the greedy user scheduling approach, can achieve the best outage performance. The max-min scheduling approach can outperform   random relaying, since its diversity gain can be improved when more users join in the competition, as shown in Theorem \ref{theorem 2}. However, it will result in some performance loss compared to Approach I and Approach III, since it cannot achieve the full diversity gain, as indicated in Theorem \ref{theorem 2}.

\begin{figure}[!htbp]\centering
    \epsfig{file=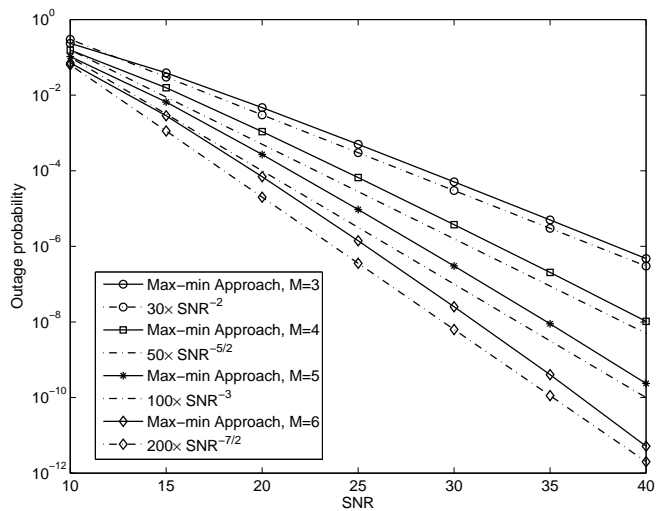, width=0.48\textwidth, clip=}
\caption{Verification of the diversity order for the max-min scheduling approach.   Only one user pair is scheduled.  $\eta=1$. The targeted data rate is $R=2$ BPCU.}\label{fig_3}\vspace{-1em}
\end{figure}

Since the main focus of this paper is to study the performance of the max-min user scheduling approach, Fig. \ref{fig_3} is provided in order to closely examine the diversity order achieved by this approach. Particularly the analytical results developed in Theorem \ref{theorm1} are used to generate the curves of outage probabilities. To clearly demonstrate achievable diversity gains, auxiliary lines with the diversity order of $\frac{M+1}{2}$  are also shown as a benchmark. As can be seen from the figure, the outage probability curves for the max-min approach are always parallel to the benchmarking curves. Recall that the diversity order is indicated by the slope of an outage probability curve. Therefore Fig. \ref{fig_3} confirms that the diversity order achieved by the max-min approach is $\frac{M+1}{2}$, as indicated by   Theorem \ref{theorem 2}. The reason for this loss of diversity gains is that the max-min approach treats the source-relay channels and the relay-destination channels equally important when user scheduling is carried out. However, when an energy harvesting relay is used, the source-relay channels become more important, since they affect not only the transmission reliability during the first phase, but also the transmission power for the second phase.

\begin{figure}[!htbp]\centering
    \epsfig{file=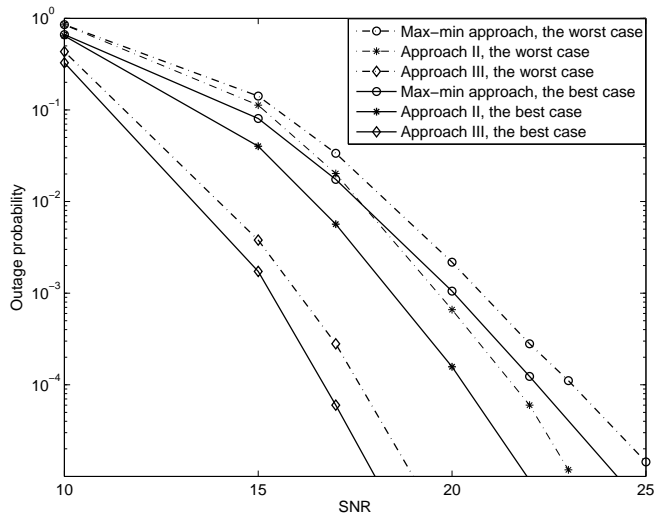, width=0.48\textwidth, clip=}
\caption{Comparison of various user scheduling approaches. The total number of user pairs is $M=10$, $\eta=1$ and two user pairs are scheduled, $m=2$.  }\label{fig_4}\vspace{-1em}
\end{figure}

\begin{figure}[!htbp]\centering
    \epsfig{file=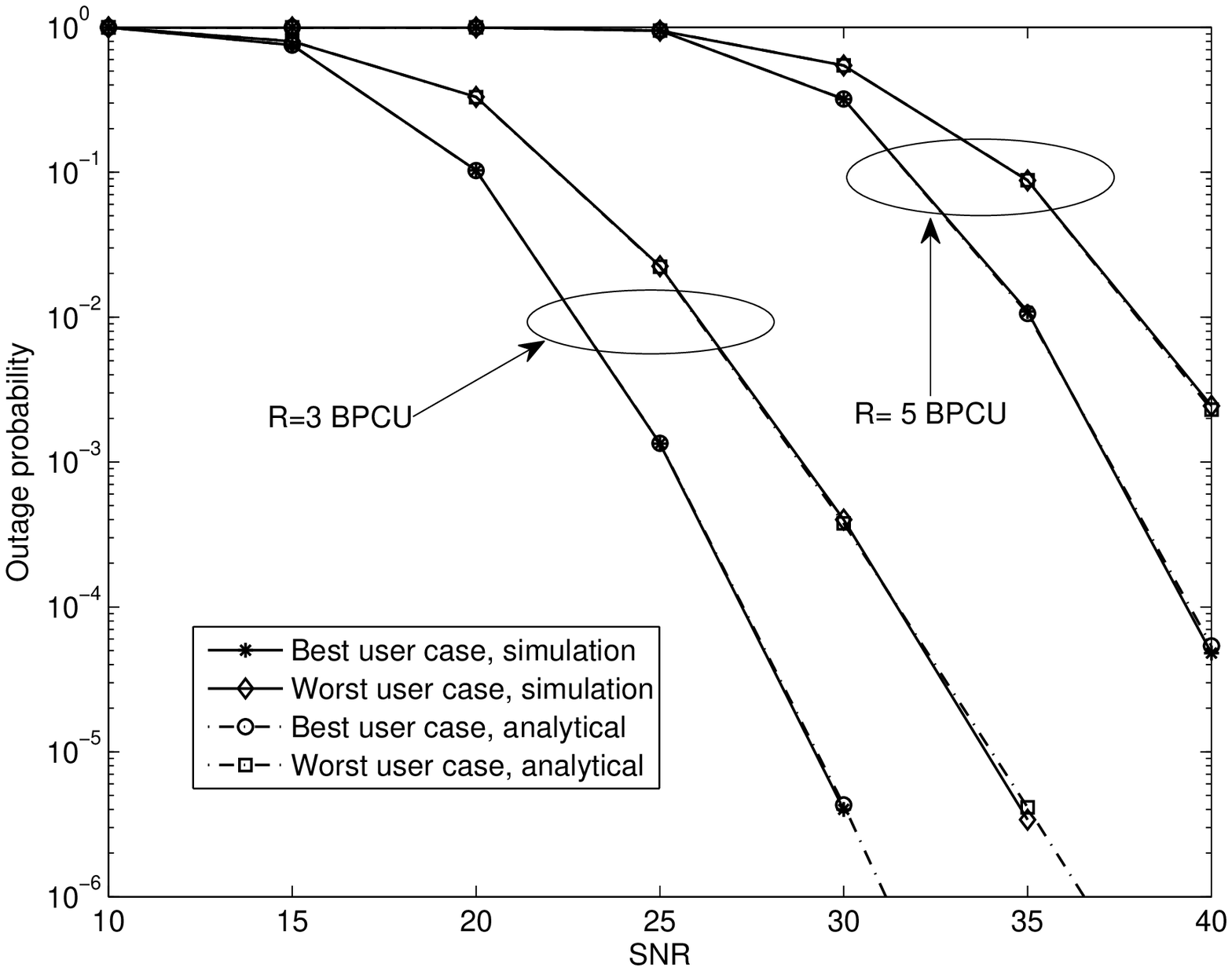, width=0.48\textwidth, clip=}
\caption{Analytical results vs computer simulations. The total number of user pairs is $M=6$, $\eta=1$ and three user pairs are scheduled, $m=3$.  }\label{fig_5}\vspace{-1em}
\end{figure}
In Figs. \ref{fig_4} and \ref{fig_5} we will focus on the scenario when multiple user pairs are scheduled. Particularly, in Fig. \ref{fig_4} we compare the outage performance achieved by the three schemes, the max-min approach and the two approaches proposed in Section \ref{section IV}. The total number of the user pairs is $M=10$ and two user pairs will be scheduled. Since the scheduled users experience different outage performance, in the figure  we show   the outage performance for the user with the strongest SNR and also the user with the weakest SNR. As can be observed from the figure, Approach III, the greedy user scheduling approach, can achieve the best outage performance, and the max-min approach achieves the worst performance. But it is worthy to point out that Approach II outperforms the max-min approach at a price of high computational complexity, since Approach II needs to enumerate all possible combinations of the user pairs. In Fig. \ref{fig_5}, we evaluate the accuracy of the analytical results  developed in Lemma \ref{lemma 2}, by comparing the outage probability calculated using \eqref{eqlemma2} to computer simulations. The total number of the user pairs is $M=6$ and three user pairs will be scheduled. As can be observed from the figure, the developed analytical results match the computer simulations exactly.

\begin{figure}[!htbp]\centering
    \epsfig{file=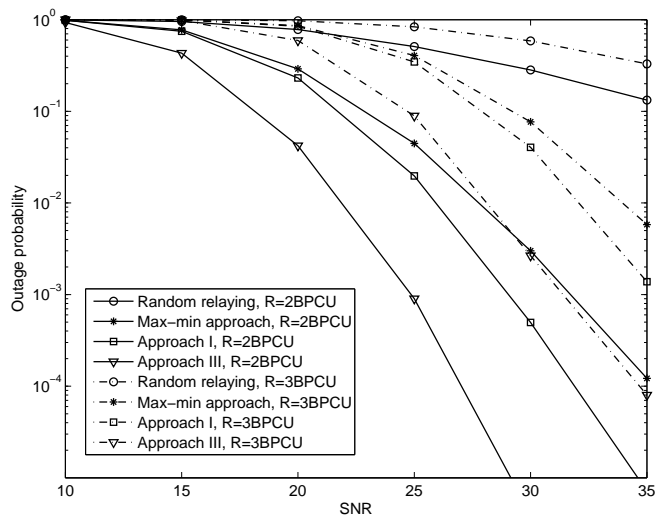, width=0.48\textwidth, clip=}
\caption{Comparison of various user scheduling approaches. $\eta=0.5$.  The total number of user pairs is $M=6$, and one user pair is scheduled, $m=1$.  }\label{fig_6}\vspace{-1em}
\end{figure}

\begin{figure}[!htbp]\centering
    \epsfig{file=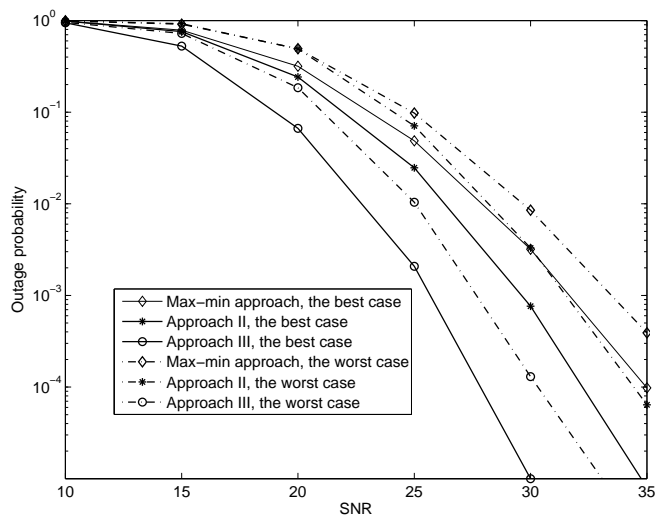, width=0.48\textwidth, clip=}
\caption{Comparison of various user scheduling approaches. $\eta=0.5$.  The total number of user pairs is $M=6$, and two user pairs are scheduled, $m=3$. The targeted data rate is $R=2$ BPCU. }\label{fig_7}\vspace{-1em}
\end{figure}

Finally we present some simulation results when $\eta<1$ and the large scale path loss is considered. Particularly consider a disk with the relay at its center and its diameter as $4$ meters. The $M$ pairs of sources and destinations are uniformly deployed in this disc, and the used path loss exponent is $2$. In Fig. \ref{fig_6} and Fig. \ref{fig_7}, the performance of the user scheduling approaches for the cases of $m=1$ and $m=2$ are shown, respectively. As can be seen from Fig. \ref{fig_6}, the use of the user scheduling approaches can improve the system performance compared to the random relaying scheme. Another observation from both figures is that, among all the opportunistic scheduling approaches,  the max-min approach achieves the worst performance,  and the greedy approach outperforms the other user scheduling approaches, which is  consistent to the previous figures.


\section{Conclusions} \label{section conclusion}
In this paper, we  considered an energy harvesting cooperative network with $M$ source-destination  pairs and one relay, where the relay   schedules only $m$ user pairs for transmissions. It is important to point out that for the special case of $m=1$, the addressed scheduling problem is the same as relay selection for the scenario with one source-destination  pair and $M$ relays. The main contribution of this paper is to show  that  the use of the max-min criterion will result in loss of diversity gains, when an energy harvesting relay is employed.  Particularly when only one user is scheduled, analytical results have been developed to demonstrate that  the diversity gain achieved by the max-min criterion is only $\frac{M+1}{2}$, much less than the maximal diversity gain $M$. Motivated by this  performance loss,   a few user scheduling approaches  tailored to energy harvesting networks have been proposed and their performance is analyzed. Simulation results have been  provided to demonstrate the accuracy of the developed analytical results and facilitate the performance comparison. When developing user scheduling approaches, only reception reliability is considered, and it is assumed that the network is delay tolerant. It is a promising future direction to study how to achieve a balanced tradeoff between reception reliability and user delay.
 \bibliographystyle{IEEEtran}
\bibliography{IEEEfull,trasfer}
\appendix
\textsl{Proof of Theorem \ref{theorm1} :}
To simplify   notation, define $x=|h_{i^*}|^2$ and $y=|g_{i^*}|^2$, and the outage probability in \eqref{outage probability} can be expressed as follows:
\begin{eqnarray}
\mathrm{P}_{o}\triangleq\mathrm{P}\left(x<\epsilon\right) + \mathrm{P}\left((x-\epsilon)y<\epsilon_1, x>\epsilon \right).
\end{eqnarray}
The scheduling strategy has changed the statistical property of $x$ and $y$, but the density function of $\min\{x,y\}$ can be found simply by applying order statistics. To use such a density function,  we need to first rewrite the outage probability as follows:
\begin{eqnarray}
\mathrm{P}_{o}&=&\mathrm{P}\left(x<\epsilon,x>y\right) + \mathrm{P}\left((x-\epsilon)y<\epsilon_1, x>\epsilon, x>y \right)\\ \nonumber &&+\mathrm{P}\left(x<\epsilon,x<y\right) + \mathrm{P}\left((x-\epsilon)y<\epsilon_1, x>\epsilon, x<y \right).
\end{eqnarray}
Converting the joint probabilities to conditional probabilities, the outage probability is given by
\begin{eqnarray}
\mathrm{P}_{o}&=&\mathrm{P}\left(x<\epsilon|x>y\right)\mathrm{P}(x>y) + \mathrm{P}\left((x-\epsilon)y<\epsilon_1, x>\epsilon| x>y \right)\mathrm{P}(x>y)\\ \nonumber &&+\mathrm{P}\left(x<\epsilon|x<y\right)\mathrm{P}(x<y) + \mathrm{P}\left((x-\epsilon)y<\epsilon_1, x>\epsilon|x<y \right)\mathrm{P}(x<y).
\end{eqnarray}
Since the incoming and outgoing channels at the relay are independent and identically distributed, we have $\mathrm{P}(x>y)=\mathrm{P}(x<y)=\frac{1}{2}$. Consequently the outage probability can   be expressed as in the following form:
\begin{eqnarray}\label{long1}
\mathrm{P}_{o}&=&\frac{1}{2}\underset{Q_1}{\underbrace{\mathcal{E}_{y|x>y}\left\{\mathrm{P}\left(x<\epsilon|x>y\right)\right\}}} + \frac{1}{2}\underset{Q_2}{\underbrace{\mathcal{E}_{y|x>y}\left\{\mathrm{P}\left((x-\epsilon)y<\epsilon_1, x>\epsilon| x>y \right)\right\}}}\\ \nonumber &&+\frac{1}{2}\underset{Q_3}{\underbrace{\mathcal{E}_{x|x<y}\left\{\mathrm{P}\left(x<\epsilon|x<y\right)\right\}}} + \frac{1}{2}\underset{Q_4}{\underbrace{\mathcal{E}_{y|x>y}\mathrm{P}\left((x-\epsilon)y<\epsilon_1, x>\epsilon |x<y \right)}},
\end{eqnarray}
where $\mathcal{E}\{\cdot\}$ denotes the expectation operation. The rationale to have the above expression is following. Take $Q_1$ as an example. $Q_1$ can be calculated in two steps. The first step is to calculate $Q_1$  by  treating $y$ as a constant and using the condition $x>y$. The second step is to calculate the expectation of the probability   by using the density function of $y$. Since $x>y$,  $y=\min\{x,y\}$, and the density function of $y$ can be found easily. In the following the four terms $Q_i$ will be evaluated individually.

\subsubsection{Calculating  $Q_1$}
We start from the calculation of $Q_1$, the first terms in \eqref{long1}. In particular, $Q_1$ can be expressed as follows:
\begin{eqnarray}\label{long2}
Q_1&=& \int^{\epsilon}_{0}\int^{\epsilon}_{y}f_{x|x>y,y}(x)dx f_{y|x>y}(y)dy,
\end{eqnarray}
where $f_{x|x>y,y}(x)$ is the pdf of $x$ conditioned on a fixed $y$ and $x>y$,  and $f_{y|x>y}(y)$ is the pdf of $y$ also conditioned on $x>y$.

To find the two conditional pdfs, we first define $x_i=|h_i|^2$ and $y_i=|g_i|^2$,   $z_i=\min\{x_i,y_i\} $, and $z=\min \{z_i, 1\leq i \leq M\}$.
From order statistics \cite{David03}, the pdf of $z_i $ is $f_{z_i}(z)=2e^{-2z}$, and the pdf of $z $ can be found as follows:
\begin{eqnarray}\label{short1}
f_{z}(z) = 2Me^{-2z}\left(1-e^{-2z}\right)^{M-1}.
\end{eqnarray}
Conditioned on $x>y$, the pdf of $y$ is the same as $z$, i.e. $f_{y|x>y}(y)=f_z(y)$. On the other hand, conditioned on a fixed $y$ and $x>y$, the cumulative distribution function (CDF) of $x$ can be found as follows:
\begin{eqnarray}\label{cox}
F_{x|x>y,y}(x) = \frac{e^{-y}-e^{-x}}{e^{-y}},
\end{eqnarray}
where the factor  $e^{-y}$ at the denominator is to ensure   $F_{x|x>y}(x)\rightarrow 1$ when $x\rightarrow \infty$.

By using the obtained conditional pdfs,   $Q_1$ can be calculated as follows:
\begin{eqnarray}
Q_1&=& \int^{\epsilon}_{0}\int^{\epsilon}_{z}f_{x|x>y}(x)dx f_{z}(z)dy\\ \nonumber &=&  e^{-\epsilon}\int^{\epsilon}_{0} \left(1-e^{-2z}  \right)^{M}e^{z}dz.
\end{eqnarray}
By applying binomial expansions, we obtain the following:
\begin{eqnarray}
Q_1  &=& e^{-\epsilon}\sum_{i=0}^{M}{M \choose i} \frac{(-1)^i}{2i-1}\left(1-e^{-(2i-1)\epsilon}\right).
\end{eqnarray}

\subsubsection{Calculating $Q_2$}
Recall that $Q_2=\mathcal{E}_{y|x>y}\left\{\mathrm{P}\left((x-\epsilon)y<\epsilon_1, x>\epsilon| x>y \right)\right\}$. The conditional density functions, $f_{x|x>y,y}(x)$ and $f_{y|x>y}(y)$, obtained in \eqref{short1} and \eqref{cox} can be used again.  An important step to calculate $Q_2$ is to determine the domain of integration.  The constrains, $x>y$ and $x<\frac{\epsilon_1}{y}+\epsilon$, imply that $y<\frac{\epsilon_1}{y}+\epsilon$. Together with the additional constraint, $x>\epsilon$, the integration domain for $Q_2$ is given by
\begin{eqnarray}
\left\{\begin{array}{ll}y<x <\frac{\epsilon_1}{y}+\epsilon, & if \quad  \epsilon<y<\epsilon_0 \\ \epsilon<x <\frac{\epsilon_1}{y}+\epsilon, & if \quad  0\leq y <\epsilon \end{array} \right.,
\end{eqnarray}
where $\epsilon_0\triangleq \frac{\epsilon +\sqrt{\epsilon^2+4\epsilon_1}}{2}$ is the positive root of $y^2-\epsilon y-\epsilon=0$, due to the constraint $y<\frac{\epsilon_1}{y}+\epsilon$.

With the obtained integration domain, $Q_2$ can be rewritten as follows:
\begin{eqnarray}
Q_2 &=& \int^{\epsilon}_{0}\int^{\epsilon+\frac{\epsilon_1}{y}}_{\epsilon}f_{x|x>y}(x)dx f_{y|x>y}(y)dy\\ \nonumber &&+ \int^{\epsilon_0}_{\epsilon}\int^{\epsilon+\frac{\epsilon_1}{y}}_{y}f_{x|x>y}(x)dx f_{y|x>y}(y)dy\\ \nonumber &=&
\underset{Q_{21}}{\underbrace{\int^{\epsilon}_{0} \left(\frac{e^{-\epsilon} - e^{-\epsilon-\frac{\epsilon_1}{y}}}{e^{-y}}\right)f_{y|x>y}(y)dy}}\\ \nonumber &&+
\underset{Q_{22}}{\underbrace{\int^{\epsilon_0}_{\epsilon} \left(\frac{e^{-y} - e^{-\epsilon-\frac{\epsilon_1}{y}}}{e^{-y}}\right)f_{y|x>y}(y)dy}}.
\end{eqnarray}
Now applying the conditional pdf of $y$, the first factor, $Q_{21}$, in the above equation can be expressed as follows:
\begin{eqnarray}\label{q21}
Q_{21} &=& 2Me^{-\epsilon} \int^{\epsilon}_{0} \left(1-e^{-\frac{\epsilon_1}{y}}\right)e^{-y}\left(1-e^{-2y}\right)^{M-1}dy
\\ \nonumber &=& 2Me^{-\epsilon} \sum^{M-1}_{i=0}{M-1 \choose i} (-1)^i  \left(\frac{1-e^{-(2i+1)\epsilon}}{2i+1} -  \int^{\epsilon}_{0}e^{-(2i+1)y-\frac{\epsilon_1}{y}} dy \right).
\end{eqnarray}
Similarly the factor $Q_{22}$ can be calculated as follows:
\begin{eqnarray}\label{q22}
Q_{22} &=& 2M  \int^{\epsilon_0}_{\epsilon} \left(e^{-y}- e^{-\epsilon-\frac{\epsilon_1}{y}}\right)e^{-y}\left(1-e^{-2y}\right)^{M-1}dy
\\ \nonumber &=& 2M  \sum^{M-1}_{i=0}{M-1 \choose i} (-1)^i  \left(\frac{e^{-(2i+2)\epsilon}-e^{-(2i+2)\epsilon_0}}{2i+2} -  e^{-\epsilon} \int^{\epsilon_0}_{\epsilon}e^{-(2i+1)y-\frac{\epsilon_1}{y}} dy \right).
\end{eqnarray}
By combining \eqref{q21} and \eqref{q22}, the factor $Q_2$ can be expressed as follows:
\begin{eqnarray}\label{q2}
Q_{2} &=& 2M \sum^{M-1}_{i=0}{M-1 \choose i} (-1)^i  \left(\frac{e^{-\epsilon}-e^{-(2i+2)\epsilon}}{2i+1} +\frac{e^{-(2i+2)\epsilon}-e^{-(2i+2)\epsilon_0}}{2i+2} \right. \\ \nonumber &&\left.- e^{-\epsilon}  \int^{\epsilon_0}_{0}e^{-(2i+1)y-\frac{\epsilon_1}{y}} dy \right).
\end{eqnarray}

\subsubsection{Calculating $Q_4$}
Recall that $Q_4=\mathcal{E}_{y|x>y}\mathrm{P}\left((x-\epsilon)y<\epsilon_1, x>\epsilon |x<y \right)$. Again it is important to determine the  integration domain of $Q_4$. Particularly, the integral constraints, $y< \frac{\epsilon_1}{x-\epsilon}$, $x>\epsilon$ and $x<y$, imply the inegration domain of $x<y< \frac{\epsilon_1}{x-\epsilon}$ and $\epsilon<x<\epsilon_0$, where the inequality of $x<\epsilon_0$ is due to $ x<\frac{\epsilon_1}{x-\epsilon}$, i.e. $x^2-\epsilon x-\epsilon_1<0$. By applying the obtained integration domain, $Q_4$  is calculated as follows:
\begin{eqnarray}
Q_4 &=& \int^{\epsilon_0}_{\epsilon}\int^{ \frac{\epsilon_1}{x-\epsilon}}_{x}f_{y|x<y,x}(y)dy f_{x|x<y}(x)dx\\ \nonumber &=&
\int^{\epsilon_0}_{\epsilon} \left(\frac{e^{-x} - e^{-\frac{\epsilon_1}{x-\epsilon}}}{e^{-x}}\right)f_{x|x<y}(x)dx,
\end{eqnarray}
where the last equation follows from the  symmetry of incoming and outgoing channels, i.e. $f_{y|x<y,x}(y)=f_{x|y>x,y}(x)$. Similarly we have $f_{x|x<y}(x)=f_{y|x>y}(y)$, which   yields the following expression of $Q_4$:
\begin{eqnarray}
Q_4 &=& 2M\int^{\epsilon_0}_{\epsilon} \left(1-e^{x-\frac{\epsilon_1}{x-\epsilon}}\right)e^{-2x} \left(1-e^{-2x}\right)^{M-1}dx \\ \nonumber
&=& 2M\sum^{M-1}_{i=0} {M-1\choose i} (-1)^i \left(\frac{e^{-2(i+1)\epsilon}-e^{-2(i+1)\epsilon_0}}{2(i+1)} -  \int^{\epsilon_0}_{\epsilon} e^{-(2i+1)x-\frac{\epsilon_1}{x-\epsilon}}dx\right).
\end{eqnarray}

 On the other hand, $Q_3$ can be easily calculated as $Q_3= F_z(\epsilon) = \left(1-e^{-2\epsilon}\right)^M $, where $F_z(z)$ is the CDF corresponding to the pdf in \eqref{short1}.
Therefore, the overall outage probability can be expressed as
\begin{eqnarray}\nonumber
\mathrm{P}_o&=& \frac{e^{-\epsilon}}{2}\sum_{i=0}^{M}{M \choose i} \frac{(-1)^i}{2i-1}\left(1-e^{-(2i-1)\epsilon}\right) \\\nonumber &&+M \sum^{M-1}_{i=0}{M-1 \choose i} (-1)^i  \left(\frac{e^{-\epsilon}-e^{-(2i+2)\epsilon}}{2i+1} +\frac{e^{-(2i+2)\epsilon}-e^{-(2i+2)\epsilon_0}}{2i+2} - e^{-\epsilon}  \int^{\epsilon_0}_{0}e^{-(2i+1)y-\frac{\epsilon_1}{y}} dy \right)
\\ \nonumber &&+  \frac{\left(1-e^{-2\epsilon}\right)^M}{2}+ M\sum^{M-1}_{i=0} {M-1\choose i} (-1)^i \left(\frac{e^{-2(i+1)\epsilon}-e^{-2(i+1)\epsilon_0}}{2(i+1)} -  e^{-(2i+1)\epsilon}\int^{\epsilon_0-\epsilon}_{0} e^{-(2i+1)x-\frac{\epsilon_1}{x }}dx\right),
\end{eqnarray}
and the proof of the theorem is completed.
 \hspace{\fill}$\blacksquare$\newline

\textsl{Proof of Theorem \ref{theorem 2} :}
To simplify the analytical development, we let $\eta=1$, which means $\epsilon_1=\epsilon$. Note that this simplification has no impact on the developed analytical results, since the diversity order is obtained at high SNR. As shown in \eqref{long1}, the outage probability can be expressed as $\mathrm{P}_o=\frac{1}{2}\sum^{4}_{l=1}Q_l$.  In the following the asymptotic study for  the four terms will be  carried out individually.

 \setcounter{subsubsection}{0}
 \subsubsection{Asymptotic study of $Q_1$} The aim of the asymptotic study is to convert $Q_1$ in a form of $t\epsilon^d$, where $t$ should be a constant, not a function of $\epsilon$, and $d$ will be used to determine the diversity order.  By applying   series expansion of exponential functions, $Q_1$, the first term in \eqref{long1}, can be expressed as follows:
\begin{eqnarray}
Q_1  &=& e^{-\epsilon}\sum_{i=0}^{M}{M \choose i} \frac{(-1)^i}{2i-1}\left(1-\sum_{k=0}^{\infty}\frac{(-1)^k (2i-1)^k \epsilon^k}{k!} \right)\\ \nonumber
&=&-e^{-\epsilon}\sum_{k=1}^{\infty}\frac{ (-1)^k  \epsilon^k}{k!}  \sum_{i=0}^{M}{M \choose i} (-1)^i (2i-1)^{k-1}.
\end{eqnarray}
Compared to the expression of $Q_1$in \eqref{theorem eq}, the above form is more complicated, but facilitates the asymptotic studies as shown in the following.
Recall the following two properties about the sums of binomial coefficients: \cite{GRADSHTEYN}
\begin{equation}\label{sum1}
\sum^{M}_{i=0}(-1)^i {M \choose i} i^{j}=0,
\end{equation}
for $0\leq j\leq (M-1)$, and
\begin{equation}\label{sum2}
\sum^{M}_{i=0}(-1)^i {M \choose i} i^{M}=(-1)^MM!.
\end{equation}
These prosperities are useful to remove the terms at the order of $\epsilon^{d}$, $d<\frac{M+1}{2}$, from $Q_1$, as described in the following.
To make the above properties applicable, we rewrite $Q_1$ as follows:
\begin{eqnarray}
Q_1
&=&-e^{-\epsilon}\sum_{k=1}^{\infty}\frac{ (-1)^k  \epsilon^k}{k!}  \sum_{i=0}^{M}{M \choose i} (-1)^i \sum^{k-1}_{j=0}{k-1 \choose j}(-1)^{k-1-j}2^ji^j.
\end{eqnarray}
All the terms with $i^j$ for $j\leq (M-1)$ can be removed because of \eqref{sum1}. At high SNR, i.e. $\epsilon\rightarrow 0$, all the factors with $\epsilon^{k}$ for $k\geq (M+2)$ can be also ignored. So the dominant factor of $Q_1$ will be the one at the order of $\epsilon^{M+1}$. By applying \eqref{sum2}, $Q_1$ can be approximated as follows:
\begin{eqnarray}
Q_1
&\approx& -\frac{ (-1)^{M+1}  \epsilon^{M+1}}{(M+1)!}  \sum_{i=0}^{M}{M \choose i} (-1)^i   2^Mi^M\\ \nonumber &=&-\frac{ (-1)^{M+1}  \epsilon^{M+1}}{(M+1)!}  2^M (-1)^MM! = \frac{2^M\epsilon^{M+1}}{M+1}.
\end{eqnarray}
Therefore the first factor of the outage probability expression in \eqref{long1} is at the order of $\epsilon^{M+1}$.

 \subsubsection{Asymptotic study of $Q_2$} The approximation of $Q_2$ is more difficult than that of $Q_1$, since $Q_2$ contains an integral which cannot be expressed analytically. As shown in \eqref{q2}, $Q_2$ can be re-written as follows:
\begin{eqnarray}\nonumber
Q_{2} &=&2M\underset{\tilde{Q}_{21}}{\underbrace{  \sum^{M-1}_{i=0}{M-1 \choose i} (-1)^i   \frac{e^{-\epsilon}-e^{-(2i+2)\epsilon}}{2i+1}}} +2M\underset{\tilde{Q}_{22}}{\underbrace{\sum^{M-1}_{i=0}{M-1 \choose i} (-1)^i \frac{e^{-(2i+2)\epsilon}-e^{-(2i+2)\epsilon_0}}{2i+2}}}   \\   && -2Me^{-\epsilon} \underset{\tilde{Q}_{23}}{\underbrace{\sum^{M-1}_{i=0}{M-1 \choose i} (-1)^i   \int^{\epsilon_0}_{0}e^{-(2i+1)y-\frac{\epsilon}{y}} dy  }}.\label{qwa2}
\end{eqnarray}
Again by applying the properties in \eqref{sum1} and \eqref{sum2}, $  \tilde{Q}_{21}$ can be approximated as follows:
\begin{eqnarray}\label{q211}
  \tilde{Q}_{21}&=&
  e^{-\epsilon}\sum^{M-1}_{i=0}{M-1 \choose i} (-1)^{i+1} \sum^{\infty}_{k=1} \frac{(-1)^k}{k!}\epsilon^k \sum^{k-1}_{j=0} {k \choose j} 2^j i^j \\ \nonumber &\approx&
 e^{-\epsilon} \frac{2^{M-1}\epsilon^M}{M}.
\end{eqnarray}
Similarly the factor $  \tilde{Q}_{22}$ can be approximated as follows:
\begin{eqnarray}\label{q222}
 \tilde{Q}_{22}   &=&\sum^{M-1}_{i=0}{M-1 \choose i} (-1)^i  \left(\sum^{\infty}_{k=1}\frac{(-1)^k}{k!}2^{k-1}\left(\epsilon^{k}-\epsilon_0^k\right) \sum^{k-1}_{j=0}{k-1 \choose j} i^j\right)
 \\ \nonumber &\approx&   \frac{(-1)^M}{M!}2^{M-1}\left(\epsilon^{M}-\epsilon_0^M\right)  (-1)^{M-1}(M-1)! = \frac{2^{M-1}\left(\epsilon_0^M-\epsilon^M\right)}{M}.
\end{eqnarray}
Different from $\tilde{Q}_{21}$ and $\tilde{Q}_{22}$, it is difficult to directly find the the closed form of the asymptotic expression for the term $\tilde{Q}_{23}$. Instead, we will first develop the upper and lower bounds on $\tilde{Q}_{23}$ and then show that they  converge at high SNR. Observe that for the integral of  $\tilde{Q}_{23}$, $ \int^{\epsilon_0}_{0}e^{-(2i+1)y-\frac{\epsilon}{y}} dy $, the range of $y$ is from $0$ to $\epsilon_0$, so $y\rightarrow 0$ at high SNR. Therefore the term in the integral, $e^{-(2i+1)y}$, can be approximated at high SNR. This observation motivates us to rewrite $\tilde{Q}_{23}$ as follows:
 \begin{eqnarray}
\tilde{Q}_{23}&=&  \sum^{M-1}_{i=0}{M-1 \choose i} (-1)^i      \int^{\epsilon_0}_{0}\left(\sum^{\infty}_{k=0} \frac{(-1)^k(2i+1)^ky^k}{k!}  \right)e^{-\frac{\epsilon}{y}} dy \\ \nonumber &=& \int^{\epsilon_0}_{0}\left(\sum^{\infty}_{k=0} \frac{(-1)^ky^k}{k!} \sum^{M-1}_{i=0}{M-1 \choose i} (-1)^i      (2i+1)^k\right)  e^{-\frac{\epsilon}{y}} dy.
\end{eqnarray}
By using the properties in \eqref{sum1} and \eqref{sum2},  $\tilde{Q}_{23}$ can be approximated as follows:
 \begin{eqnarray}
\tilde{Q}_{23} &\approx&\int^{\epsilon_0}_{0}\left(  \frac{(-1)^{M-1}y^{M-1}}{(M-1)!}   2^{M-1} (-1)^{M-1}(M-1)!   \right)  e^{-\frac{\epsilon}{y}} dy\\ \nonumber&=& \int^{\epsilon_0}_{0} 2^{M-1} y^{M-1}e^{-\frac{\epsilon}{y}} dy\triangleq \bar{Q}_{23},
\end{eqnarray}
where the approximation follows from the fact that $0\leq y\leq \epsilon_o$ and $\epsilon_0\rightarrow 0$ at high SNR.  To obtain the upper and lower bounds on $\bar{Q}_{23}$, the use of the inequalities for exponential functions yields the following:
 $$1-\frac{\epsilon}{y}\leq e^{-\frac{\epsilon}{y}}\leq \frac{1}{1+\frac{\epsilon}{y}},$$ for $0\leq y\leq \epsilon_0$.
Now the upper bound of $Q_{23}$ can be computed as follows:
 \begin{eqnarray}
 \bar{Q}_{23}  &\leq &2^{M-1}\int^{\epsilon_0}_{0}  y^{M-1}\frac{1}{1+\frac{\epsilon}{y}} dy
  \\ \nonumber &= &2^{M-1}\sum^{M}_{i=1}{M \choose i} (-1)^{M-i}\epsilon^{M-i}\frac{(\epsilon_0+\epsilon)^i-\epsilon^i}{i}  +2^{M-1}(-1)^M\epsilon^M\ln \frac{\epsilon_0+\epsilon}{\epsilon}.
\end{eqnarray}
At high SNR, $\epsilon\rightarrow 0$, and $\epsilon_0\rightarrow \epsilon^{\frac{1}{2}}$. Therefore the dominant factors in the upper bound of $\bar{Q}_{23}$ are the terms with $i=M$ and $i=M-1$, which means
 \begin{eqnarray}\label{q233}
 \bar{Q}_{23} &\leq &2^{M-1} \left( \frac{(\epsilon_0+\epsilon)^M-\epsilon^M}{M}  -\epsilon \frac{M(\epsilon_0+\epsilon)^{M-1}-M\epsilon^{M-1}}{M-1} \right).
\end{eqnarray}
Combining \eqref{q211}, \eqref{q222} and \eqref{q233}, $Q_2$ can be lower bounded as follows:
%
\begin{eqnarray}\label{lower bound}
Q_2 &=& 2M\tilde{Q}_{21}+2M\tilde{Q}_{22} -2Me^{-\epsilon}\tilde{Q}_{23}\\ \nonumber &\geq &  2^{M}\epsilon^M+  2^{M}\left(\epsilon_0^M-\epsilon^M\right)-  2^{M}  (\epsilon_0+\epsilon)^M+2^M\epsilon^M\\ \nonumber &&+2^{M} \epsilon \frac{M^2(\epsilon_0+\epsilon)^{M-1}}{M-1}-2^{M} \epsilon \frac{M^2\epsilon^{M-1}}{M-1}\\ \nonumber &\underset{(a)}{\approx}&     2^M\left(\epsilon_0^M - \epsilon_0^M-M\epsilon\epsilon_0^{M-1} +\epsilon \frac{M^2\epsilon_0^{M-1}}{M-1} \right) \\ \nonumber &=& \frac{2^M}{M-1}\epsilon \epsilon_0^{M-1}\rightarrow \epsilon^{\frac{M+1}{2}},
\end{eqnarray}
where $(a)$ is obtained by keeping only the terms at the order of $\epsilon_0^M$ and $\epsilon\epsilon_0^{M-1}$.

The lower bound of $\bar{Q}_{23}$ can be obtained as follows:
 \begin{eqnarray}\label{eqxxx}
 \bar{Q}_{23}   &\geq &2^{M-1}\int^{\epsilon_0}_{0}  y^{M-1}\left(1-\frac{\epsilon}{y}\right) dy
 \\ \nonumber &= &2^{M-1}\left(\frac{\epsilon_0^M}{M}-\epsilon \frac{\epsilon_0^{M-1}}{M-1}\right).
\end{eqnarray}
Combining \eqref{eqxxx} with \eqref{q211} and \eqref{q222}, the upper bound of  $Q_2$ can be asymptotically shown in the following:
\begin{eqnarray}\label{upper bound}
Q_2 &=& 2M\tilde{Q}_{21}+2M\tilde{Q}_{22} -2Me^{-\epsilon}\tilde{Q}_{23}\\ \nonumber &\leq &  2^{M}\epsilon^M+  2^{M}\left(\epsilon_0^M-\epsilon^M\right)-  2^{M}  \left(\epsilon_0^M -\epsilon \frac{M\epsilon_0^{M-1}}{M-1}\right)\\ \nonumber &=& \frac{M2^M}{M-1}\epsilon \epsilon_0^{M-1}\rightarrow \epsilon^{\frac{M+1}{2}}.
\end{eqnarray}
As can be observed from \eqref{upper bound} and \eqref{lower bound}, the upper and lower bounds converge at high SNR, which implies
\begin{eqnarray}
Q_2 \rightarrow \epsilon^{\frac{M+1}{2}}.
\end{eqnarray}

\subsubsection{Asymptotic study of $Q_4$} First rewrite $Q_4$ in the following expression:
\begin{eqnarray}\label{qwa4}
Q_4
&=&2M\underset{\tilde{Q}_{41}}{\underbrace{\sum^{M-1}_{i=0} {M-1\choose i} (-1)^i \frac{e^{-2(i+1)\epsilon}-e^{-2(i+1)\epsilon_0}}{2(i+1)} }}\\ \nonumber &&- 2M\underset{\tilde{Q}_{42}}{\underbrace{\sum^{M-1}_{i=0} {M-1\choose i} (-1)^i  e^{-(2i+1)\epsilon}\int^{\epsilon_0-\epsilon}_{0} e^{-(2i+1)x-\frac{\epsilon}{x }}dx }}.
\end{eqnarray}
Comparing \eqref{qwa4} to \eqref{qwa2}, we observe that $\tilde{Q}_{41} $ is the same as $\tilde{Q}_{22} $, and therefore can be approximated similarly as follows:
\begin{eqnarray}
\tilde{Q}_{41} = \tilde{Q}_{22} \approx  \frac{2^{M-1}\left(\epsilon_0^M-\epsilon^M\right)}{M}.
\end{eqnarray}
Similar to $\tilde{Q}_{23} $, the term   $\tilde{Q}_{42}$ also contains an integral whose analytical closed-form expression cannot be found. Following the previous steps, we can first use the series expansion of $e^{-(2i+1)(\epsilon+x)}$ to get the following:
\begin{eqnarray}\label{qwa42}
\tilde{Q}_{42}
&=& \int^{\epsilon_0-\epsilon}_{0} \sum^{M-1}_{i=0} {M-1\choose i} (-1)^i  \sum^{\infty}_{k=0}\frac{(-1)^k}{k!} (2i+1)^k(\epsilon+x)^k e^{ -\frac{\epsilon}{x }}dx\\ \nonumber
&=& \int^{\epsilon_0-\epsilon}_{0}  \sum^{\infty}_{k=0}\frac{(-1)^k}{k!} \sum^{M-1}_{i=0} {M-1\choose i} (-1)^i \left(\sum^{k}_{j=0}{k \choose j} 2^j i^j\right)(\epsilon+x)^k e^{ -\frac{\epsilon}{x }}dx.
\end{eqnarray}
And by using   the properties in \eqref{sum1} and \eqref{sum2}, we obtain
\begin{eqnarray}
\tilde{Q}_{42}&\approx& \int^{\epsilon_0-\epsilon}_{0}   \frac{(-1)^{M-1}}{(M-1)!}  \left( 2^{M-1} (-1)^{M-1}(M-1)!\right)(\epsilon+x)^{M-1} e^{ -\frac{\epsilon}{x }}dx
\\ \nonumber
&=& \int^{\epsilon_0-\epsilon}_{0}     2^{M-1}  (\epsilon+x)^{M-1} e^{ -\frac{\epsilon}{x }}dx\triangleq \bar{Q}_{42}.
\end{eqnarray}
Again applying the upper  bound of exponential functions, we have
 \begin{eqnarray}\label{qwa43}
\bar{Q}_{42}
&\leq& \int^{\epsilon_0-\epsilon}_{0}     2^{M-1}  (\epsilon+x)^{M-1} \frac{1}{1+\frac{\epsilon}{x }}dx\\ \nonumber &=&
2^{M-1} \sum^{M-2}_{i=0} {M-2 \choose i} \epsilon^{M-2-i}\frac{(\epsilon_0-\epsilon)^{i+2}}{i+2} \\ \nonumber &\approx& 2^{M-1}\frac{(\epsilon_0-\epsilon)^{M}}{M}\rightarrow \epsilon^{-\frac{M+1}{2}}.
\end{eqnarray}
By subsisting this upper bound to the expression of $Q_4$, the lower bound of $Q_4$ is given by
 \begin{eqnarray}\label{qwa44}
Q_4
&\geq&  2^{M}\left(\epsilon_0^M-\epsilon^M\right)  - 2^{M} (\epsilon_0-\epsilon)^{M} \approx 2^MM \epsilon\epsilon_0^{M-1}.
\end{eqnarray}
On the other hand, the lower bound of $\tilde{Q}_{42}$ can be expressed as follows:
 \begin{eqnarray}\label{qwa45}
\bar{Q}_{42}
&\geq& \int^{\epsilon_0-\epsilon}_{0}     2^{M-1} x^{M-1} \left(1-\frac{\epsilon}{x }\right)dx  \\ \nonumber &=&    2^{M-1} \left(\frac{(\epsilon_0-\epsilon)^M}{M}  -\epsilon \frac{(\epsilon_0-\epsilon)^{M-1}}{M-1} \right).
\end{eqnarray}
Therefore the upper bound of $Q_4$ can be shown as follows:
 \begin{eqnarray}\label{qwa46}
Q_4
&\leq &  2^{M}\left(\epsilon_0^M-\epsilon^M\right)  - 2^{M} \left( (\epsilon_0-\epsilon)^M   -\epsilon \frac{M(\epsilon_0-\epsilon)^{M-1}}{M-1} \right) \\ \nonumber &\approx& 2^{M} \epsilon_0^M  - 2^{M} \left( \epsilon_0^M-M\epsilon\epsilon_0^{M-1}    - \frac{M\epsilon\epsilon_0^{M-1}}{M-1} \right) \\\nonumber &=& \frac{M^22^M}{M-1}\epsilon \epsilon_0^{M-1}\rightarrow \epsilon^{\frac{M+1}{2}},
\end{eqnarray}
where the approximation is carried out by keeping only the terms at $\epsilon_0^M$ and $\epsilon_0^{M-1}\epsilon$. Combining \eqref{qwa43} and \eqref{qwa46}, one can observe that the upper and lower bounds converge at high SNR, and the following conclusion can be obtained
\begin{eqnarray}
Q_4 \rightarrow \epsilon^{\frac{M+1}{2}}.
\end{eqnarray}
Applying the series expansion of exponential functions,   $Q_3$ can be simply approximated as
$
Q_3 \approx 2^M\epsilon^M
$. Therefore the asymptotic expression for the overall outage probability can be obtained as follows:
\begin{eqnarray}
\mathrm{P}_o &=& \frac{1}{2}\sum^{4}_{i=1}Q_4 \\ \nonumber &\rightarrow& \epsilon^{M+1} +2\epsilon^{\frac{M+1}{2}}+\epsilon^M\rightarrow \epsilon^{\frac{M+1}{2}},
\end{eqnarray}
and the proof of the theorem is completed.
 \hspace{\fill}$\blacksquare$\newline

\textsl{Proof of Lemma \ref{lemma 2} :}
Based on the equal power allocation strategy, the power allocated to each destination is given by $$ \left\{\begin{array}{ll}  \frac{1}{m}\sum^{m}_{i=1}P\eta\left(x_{(M-i+1)}-\epsilon\right) & if \quad M\geq n\geq m \\\frac{1}{n}\sum^{n}_{i=1}P\eta\left(x_{(M-i+1)}-\epsilon\right) & if \quad 1\leq n\leq  m-1  \end{array}\right..$$
Therefore the overall outage probability will be
\begin{eqnarray}\label{po1}
\mathrm{P}_{oi}&\triangleq& \mathrm{P}\left(|\mathcal{S}|=0\right) +\sum^{m}_{n=1}\underset{T_2}{\underbrace{\mathrm{P}\left(\left.|g_{(M-i+1)}|^2\sum^{n}_{j=1} \left(x_{(M-j+1)}-\epsilon\right) <n\epsilon_1\right||\mathcal{S}|=n\right)}} \mathrm{P}\left(|\mathcal{S}|=n\right) \\ \nonumber &&+\sum^{M}_{n=m+1}\underset{T_3}{\underbrace{\mathrm{P}\left(\left.|g_{(M-i+1)}|^2\sum^{m}_{j=1} \left(x_{(M-j+1)}-\epsilon\right) <m\epsilon_1\right||\mathcal{S}|=n\right)}} \mathrm{P}\left(|\mathcal{S}|=n\right).
\end{eqnarray}

$T_3$ can be first rewritten  as follows:
\begin{eqnarray}\label{alpham}
T_3 &=& \mathrm{P}\left(\left.y_{(M-i+1)}\alpha_m <m\epsilon_1\right||\mathcal{S}|=n\right),
\end{eqnarray}
where $\alpha_m=\sum^{m}_{j=1} \left(x_{(M-j+1)}-\epsilon\right)$ and   $m\leq n\leq M$. The condition of $T_3$ implies that there are   $n$, $n> m$, sources whose information can be decoded by the relay and   $m$ of the $n$ users will be scheduled. Therefore the conditional pdf of  $\alpha_m$ will be the same as that of $\sum^{m}_{j=1} \left(\tilde{x}_{(n-i+1)}-\epsilon\right)$, where $\tilde{x}_{(i)}$ are from the parents $\tilde{x}_i$,  and $\tilde{x}_i$, $1\leq i \leq n$, are   i.i.d. exponentially variables with the constraint $\tilde{x}_i>\epsilon$. It is straightforward to verify that  the  CDF of $\tilde{x}_i$ conditioned on $\tilde{x}_i>\epsilon$ is $ F_{\tilde{x}_i}(x) = \frac{e^{-\epsilon}-e^{-x}}{e^{-\epsilon}}$. Consequently $w_i\triangleq (\tilde{x}_i-\epsilon)$ is simply another  exponential variable. Therefore the pdf of  $\alpha_m$ is the same as the pdf of $w\triangleq\sum^{m}_{j=1}w_{(n-j+1)}$, the sum of $m$ largest order statistics chosen from $n$ i.i.d exponential variables. Following  the steps in \cite{Alam791,Dingpoor1311}, the pdf of $w$ is given by
\begin{eqnarray}
f_{w}(w) =  \sum^{n-m-1}_{k=0} d_{m,k}\left(\sum^{m}_{j=1}\frac{a_{j,k}e^{-w}w^{j-1}}{(j-1)!} +b_k e^{-\left(1+\frac{k+1}{m}\right)w}\right).
\end{eqnarray}
 From \cite{David03}, the pdf of $y_{(M-i+1)}$ is $f_{y_{(M-i+1)}}(y) = \frac{M!}{(M-i)!(i-1)!}e^{-iy} \left(1-e^{-y}\right)^{M-i}$.
So $T_3$ can be calculated as follows:
\begin{eqnarray}\label{po2}
T_3 &=&\int^{\infty}_{0} f_w(w) \int^{\frac{m\epsilon_1}{w}}_{0} f_{y_{(M-i+1)}}(y)dydw
\\ \nonumber &=&
\frac{M!}{(M-i)!(i-1)!}\sum^{M-i}_{l=0}{M-i \choose l} \frac{(-1)^l}{l+i}\int^{\infty}_{0} f_w(w) \left( 1-e^{-\frac{m\epsilon_1(l+i)}{w}}  \right)dw
\\ \nonumber &=&
\frac{M!}{(M-i)!(i-1)!}\sum^{M-i}_{l=0}{M-i \choose l} \frac{(-1)^l}{l+i} \left( 1-\underset{T4}{\underbrace{\int^{\infty}_{0} f_w(w)e^{-\frac{m\epsilon_1(l+i)}{w}}  dw}}\right).
\end{eqnarray}
The integral in the above equation can be calculated as follows:
\begin{eqnarray}
T_4&=&  \sum^{n-m-1}_{k=0} d_{m,k}\left(\sum^{m}_{j=1}\frac{a_{j,k}\int^{\infty}_{0}e^{-w}w^{j-1}e^{-\frac{m\epsilon_1(l+i)}{w}}dw}{(j-1)!}   +b_k \int^{\infty}_{0} e^{-\left(1+\frac{k+1}{m}\right)w}e^{-\frac{m\epsilon_1(l+i)}{w}}  dw \right).
\end{eqnarray}
With some straightforward manipulations, $T_4$ can be further simplified  as shown in the lemma.

$T_2$ can be first recalculated as follows:
\begin{eqnarray}
T_2 &=& \mathrm{P}\left(\left.|g_{(M-i+1)}|^2\alpha_n <n\epsilon_1\right||\mathcal{S}|=n\right),
\end{eqnarray}
where $\alpha_n=\sum^{n}_{j=1} \left(x_{(M-j+1)}-\epsilon\right)$. Different to $\alpha_m$ in \eqref{alpham}, the pdf of $\alpha_n$ can be found simply as in the following.  The condition of $T_2$ implies that there are   $n$ sources whose information can be decoded by the relay and all these users will be scheduled. Therefore the conditional pdf of  $\alpha_n$ will be the same as that of $\sum^{n}_{j=1} \left(\tilde{x}_i-\epsilon\right)$. Following the same arguments  as previously,   $(\tilde{x}_i-\epsilon)$ is simply an exponential variable, which means $\alpha_n$ is Chi-square distributed, i.e. $f_{\alpha_n}(z) = \frac{e^{-x}x^{n-1}}{(n-1)!}$.
Therefore  $T_2$ can be calculated as follows:
\begin{eqnarray}\label{po3}
T_2 &=& \int^{\infty}_{0}\frac{e^{-z}z^{n-1}}{(n-1)!}\int^{\frac{n\epsilon_1}{z}}_{0} \frac{M!}{(M-i)!(i-1)!}e^{-ix} \left(1-e^{-x}\right)^{M-i}dydz
  \\ \nonumber &=&
 \frac{M!}{(M-i)!(i-1)!(n-1)!}\sum^{M-i}_{k=0} {M-i \choose k} \frac{ (-1)^k}{k+i} \int^{\infty}_{0} \left(e^{-z}z^{n-1}-  z^{n-1}e^{-z-\frac{(k+i)n\epsilon_1}{z}} \right) dz.
\end{eqnarray}
Combining \eqref{po1}, \eqref{po2} and \eqref{po3}, and  also with some algebraic manipulation, the outage probability shown in the lemma can be obtained. The proof of the lemma is completed.
 \hspace{\fill}$\blacksquare$\newline

\textsl{Proof of Lemma \ref{lemma 3} :}
When   $m=1$,  the overall outage probability can be simplified as follows:
\begin{eqnarray}\label{po4}
\mathrm{P}_{oi}&\triangleq& \mathrm{P}\left(|\mathcal{S}|=0\right) +\sum^{M}_{n=1}T_3\mathrm{P}\left(|\mathcal{S}|=n\right) .
\end{eqnarray}
The condition that only one user pair will be scheduled can also help to simplify the expression of $T_3$   as follows:
\begin{eqnarray}
T_3&=&n\int^{\infty}_{0}  e^{-y} \left(1-e^{-y}\right)^{n-1}\int^{\frac{\epsilon_1}{y}}_{0}  d\left(1-e^{-z}\right)^{M}dy
\\ \nonumber &=&n\sum^{M}_{k=0} {M \choose k} (-1)^k \sum^{n-1}_{i=0} {n-1 \choose i} (-1)^i 2 \sqrt{\frac{k\epsilon_1}{i+1}} \mathbf{K}_1\left(2\sqrt{(i+1)k\epsilon_1}\right),
\end{eqnarray}
where the first equation follows from the density function of the largest order statistics.
Recall the series representation of  the Bessel function as follows:
\begin{eqnarray}\label{approximation2}
x\mathbf{K}_1(x) &=&1+ x \mathbf{I}_1(x)\left(\ln \frac{x}{2} +\mathbf{C}\right) -\frac{1}{2} \sum^{\infty}_{l=0}\frac{\left(\frac{x}{2}\right)^{2l+1}x}{l!(l+2)!}\left( \sum^{l}_{k=1}\frac{1}{k} +\sum^{l+2}_{k=1}\frac{1}{k}  \right)\\ \nonumber &\approx & 1+\sum^{\infty}_{q=1}\kappa_qx^{2q}\ln x,
\end{eqnarray}
 for $x\rightarrow 0$, where $\kappa_q$ is the  constant coefficient associated to $x^{2q}\ln x$. Note that the terms of $x^{2q}$ have been ignored since they are dominated by the terms of $x^{2q}\ln x$ when $x\rightarrow 0$.   It is also worthy to point out  that the exact value of $\kappa_q$ has no effect to   diversity gains. By applying the  above approximation,  we can  rewrite $T_3$ as follows:
\begin{eqnarray}\nonumber
T_3 &\approx&
n\sum^{M}_{k=0} {M \choose k} (-1)^k \sum^{n-1}_{i=0} {n-1 \choose i} \frac{(-1)^i}{i+1} \left(1+  \sum^{\infty}_{q=1} \frac{\kappa_q}{2 }\phi_{i,k}^{q}\ln \phi_{i,k} \right),
\end{eqnarray}
where $\phi_{i,k}=4 (i+1)k\epsilon_1$. We first focus on the case of $n=M$.
Since $\sum^{M-1}_{k=0}{M-1 \choose k} (-1)^k =0$, we have
\begin{eqnarray}\nonumber
M\sum^{M}_{k=0} {M \choose k} (-1)^k \sum^{M-1}_{i=0} {M-1 \choose i} \frac{(-1)^i}{i+1} \cdot 1 =0.
\end{eqnarray}
To show that the terms at the order of $\epsilon^q \ln \epsilon$, $1\leq q\leq (M-1)$, are zero, we first observe the following:
\begin{eqnarray}
\phi_{i,k}^{q}\ln \phi_{i,k}&=&4^q (i+1)^qk^q\epsilon_1^q \ln \left[4 (i+1)k\epsilon_1\right]\\ \nonumber &=&
  \underset{T_5}{\underbrace{ 4^q (i+1)^qk^q\epsilon_1^q \ln \left[4 (i+1)\epsilon_1\right]}} +\underset{T_6}{\underbrace{ 4^q (i+1)^qk^q\epsilon_1^q \ln  k }}.
\end{eqnarray}
By using the above separated expression, we can show that
\begin{eqnarray}
M\sum^{M}_{k=0} {M \choose k} (-1)^k \sum^{M-1}_{i=0} {M-1 \choose i} \frac{(-1)^i}{i+1}T_5=0,
\end{eqnarray}
since $\sum^{M}_{k=0}{M \choose k} (-1)^k k^q =0$, $1\leq q\leq (M-1)$,
and
\begin{eqnarray}
M\sum^{M}_{k=0} {M \choose k} (-1)^k \sum^{M-1}_{i=0} {M-1 \choose i} \frac{(-1)^i}{i+1}T_6=0,
\end{eqnarray}
since $\sum^{M-1}_{i=0}{M-1 \choose i} (-1)^i i^{q-1} =0$, $1\leq q\leq (M-1)$.
Therefore the term at the order of $\epsilon^{M-1} \ln \epsilon$ will be removed from $T_3$, and the overall outage probability can be expressed as
\begin{eqnarray}\nonumber
T_3  &\approx&
M\sum^{M}_{k=0} {M \choose k} (-1)^k \sum^{M-1}_{i=0} {M-1 \choose i} \frac{(-1)^i}{i+1} \left(1+  \sum^{\infty}_{q=M} \frac{\kappa_q}{2 }\phi_{i,k}^{q}\ln \phi_{i,k} \right).
\end{eqnarray}
Therefore the dominant factor is at the order of $\epsilon^M\ln \epsilon$. Similarly the dominant factors for $T_3$, $1\leq n<M$, is $\epsilon^n \ln \epsilon$. Substituting this result into \eqref{po4} and also using the fact that $\mathrm{P}\left(|\mathcal{S}|=n\right)\rightarrow \epsilon^{M-n}$,    the diversity gain of the overall outage probability will be $M$. And the proof of the lemma is completed.
 \hspace{\fill}$\blacksquare$\newline

\end{document}